\documentclass[preprintnumbers, floatfix, letterpaper, twocolumn,aps,prd,epsfig,nofootinbib,natbib,longbibliography]{revtex4-2}

\usepackage{graphicx}
\usepackage{epstopdf}
\usepackage{natbib}
\usepackage{latexsym}
\usepackage{amssymb}
\usepackage{amsmath}
\usepackage{color}
\usepackage{mathrsfs}
\usepackage{xparse}
\usepackage{bbding}
\usepackage{enumitem}
\usepackage{pifont}
\usepackage[center]{subfigure}
\usepackage[
            pdfstartview=FitH,
            bookmarksnumbered=true,
            bookmarksopen=true,
            colorlinks,
            linkcolor=blue,
            anchorcolor=green,
            citecolor=blue
            ]{hyperref}

            \usepackage{setspace}
\usepackage{amssymb}
\usepackage{graphicx,wrapfig}
\usepackage{enumerate}
\usepackage{color}
\usepackage{mathrsfs}
\usepackage{subfigure}
\usepackage{comment}
\usepackage{multirow}
\definecolor{mg}{rgb}{0.0, 0.5, 0.0}

\begin{document}
  \renewcommand\arraystretch{2}
 \newcommand{\bq}{\begin{equation}}
 \newcommand{\eq}{\end{equation}}
 \newcommand{\bqn}{\begin{eqnarray}}
 \newcommand{\eqn}{\end{eqnarray}}
 \newcommand{\nb}{\nonumber}
 \newcommand{\lb}{\label}

\newcommand{\La}{\Lambda}
\newcommand{\va}{\scriptscriptstyle}
\newcommand{\be}{\nopagebreak[3]\begin{equation}}
\newcommand{\ee}{\end{equation}}

\newcommand{\ba}{\nopagebreak[3]\begin{eqnarray}}
\newcommand{\ea}{\end{eqnarray}}

\newcommand{\la}{\label}
\newcommand{\n}{\nonumber}
\newcommand{\su}{\mathfrak{su}}
\newcommand{\SU}{\mathrm{SU}}
\newcommand{\U}{\mathrm{U}}

\def\be{\nopagebreak[3]\begin{equation}}
\def\ee{\end{equation}}
\def\ba{\nopagebreak[3]\begin{eqnarray}}
\def\ea{\end{eqnarray}}
\newcommand{\f}{\frac}
\def\rmd{\rm d}
\def\pl{\ell_{\rm Pl}}
\def\d{{\rm d}}
\def\fe{\mathring{e}^{\,i}_a}
\def\fw{\mathring{\omega}^{\,a}_i}
\def\fq{\mathring{q}_{ab}}
\def\t{\tilde}

\def\db{\delta_b}
\def\dc{\delta_c}
\def\T{\mathcal{T}}
\def\GammaE{\Gamma_{\rm ext}}
\def\GammaEb{\bar\Gamma_{\rm ext}}
\def\GammaEh{\hat\Gamma_{\rm ext}}
\def\Hee{H_{\rm eff}^{\rm ext}}
\def\H{\mathcal{H}}

\newcommand{\R}{\mathbb{R}}

 \newcommand{\cb}{\color{blue}}
    \newcommand{\cc}{\color{cyan}}
        \newcommand{\cm}{\color{magenta}}
\newcommand{\rc}{\rho^{\scriptscriptstyle{\mathrm{I}}}_c}
\newcommand{\rd}{\rho^{\scriptscriptstyle{\mathrm{II}}}_c}
\NewDocumentCommand{\evalat}{sO{\big}mm}{%
  \IfBooleanTF{#1}
   {\mleft. #3 \mright|_{#4}}
   {#3#2|_{#4}}%
}
\newcommand{\PRL}{Phys. Rev. Lett.}
\newcommand{\PL}{Phys. Lett.}
\newcommand{\PR}{Phys. Rev.}
\newcommand{\CQG}{Class. Quantum Grav.}

\title{Revisiting quantum black holes from effective loop quantum gravity}

\author{Geeth Ongole$^{a}$}
\email{Geeth$\_$Ongole1@baylor.edu}
\author{Parampreet Singh$^{b}$}
\email{psingh@lsu.edu}
\author{Anzhong Wang$^{a}$}
\email{Anzhong$\_$Wang@baylor.edu}

\affiliation{$^{a}$ GCAP-CASPER, Physics Department, Baylor University, Waco, Texas, 76798-7316, USA\\
$^{b}$ Department of Physics and Astronomy, Louisiana State University, Baton Rouge, Louisiana 70803, USA}

\begin{abstract}
We systematically study a family of loop quantizations for the classical Kruskal spacetimes using the effective description motivated from loop quantum gravity for four generic parameters, $c_o, m, \delta_b$, and $\delta_c$, where the latter two denote the  polymerization parameters that capture the underlying quantum geometry. We focus on the family where polymerization parameters are constant on dynamical trajectories and of which the Ashtekar-Olmedo-Singh (AOS) and Corichi-Singh (CS) models appear as special cases. We study general features of singularity resolution in all these models due to  quantum gravity effects and analytically extend the solutions across the white hole (WH) and black hole (BH) horizons to the exterior.  We find that the leading term in the asymptotic expansion of the Kretschmann scalar is $r^{-4}$. However, for AOS and CS models,  black holes with masses greater than solar mass, the dominant term behaves as $r^{-6}$ for the size of the observable Universe and our analysis can be used to phenomenologically constrain the choice of parameters for other models.  In addition, one can uniquely fix the parameter $c_o$ by requiring that the Hawking temperature at the BH horizon to the leading order be consistent with its classical value for a macroscopic BH.  Assuming that the BH and WH masses are of the same order, we are able to identify a family of choices of $\delta_b$ and $\delta_c$ which share all the desired properties of the AOS model.
\end{abstract}

\maketitle

\section{Introduction} \label{intro}
\renewcommand{\theequation}{1.\arabic{equation}}
\setcounter{equation}{0}

Black holes (BHs) initially were only discovered as solutions to Einstein's general relativity (GR) \cite{Schwarzschild:1916uq} but now observations have confirmed their existence \cite{LIGOScientific:2016aoc,EventHorizonTelescope:2019dse}. However, to understand these objects in entirety, GR is  insufficient as it predicts its own demise near the singularities present inside  BHs \cite{Hawking:1973uf}. Spacetimes near these singularities enter the Planckian regime  and it is widely believed that a quantum theory of gravity is necessary to understand these esoteric objects.

Loop quantum gravity is a nonperturbative theory of quantum gravity \cite{Ashtekar:2004eh}. Its techniques have been used to resolve the cosmological singularities, which is now a well-established research field, usually referred to as loop quantum cosmology (LQC)
\cite{Ashtekar:2011ni,Li:2023dwy,Agullo:2023rqq}. Since the interior of a static BH is isometric to the vacuum Kantowski-Sachs (KS) cosmological spacetime, techniques from LQC have been borrowed to study loop quantum black holes (LQBHs) \cite{Ashtekar:2023cod}. 
A majority of  LQBH models describe the spacetimes in an effective or semiclassical fashion, as such a description has been proved very successful in LQC
\cite{Singh:2012zc,Diener:2014mia,Diener:2017lde}, and captures the leading-order quantum corrections with the introduction of polymerization parameters. In particular, it shows explicitly that the big bang singularity is replaced by a  quantum bounce (see, e.g., \cite{Ashtekar:2006rx,Ashtekar:2006wn,Ashtekar:2006es}) and the subsequent cosmological perturbations are consistent with current observations (see, e.g., \cite{Li:2021mop, Agullo:2023rqq}). 
It is the general belief in the community that such an effective description should be also applicable to LQBHs, although a detailed derivation of the effective description from the full quantum dynamics of LQBHs is still absent. In this manuscript our analysis will assume the validity of this effective spacetime description.

In the quantum theory of LQBHs, the elementary variables are holonomies of  connection variables that contribute to trigonometric functions of the connection at the level of the Hamiltonian. Different quantizations of LQBH models differ in their choices of the two polymerization parameters, $\delta_b$ and $\delta_c$, which capture the edge length of the loop in the quantization procedure. At an effective level this amounts to the following replacements of the two symmetry reduced Ashtekar-Barbero connection variables $b$ and $c$ \cite{Gambini:2022hxr,Ashtekar:2023cod}:
\bqn
\lb{eq1.1}
b \rightarrow \frac{\sin(\delta_b b)}{\delta_b}, \quad c \rightarrow \frac{\sin(\delta_c c)}{\delta_c},
\eqn
in the classical Hamiltonian. It is clear that when $\delta_i \rightarrow 0\; (i = b, c)$, the classical Hamiltonian is recovered and quantum effects are expected to be negligibly small. On the other hand, such effects will become large when they are large.  Generally speaking,  $\delta_b$ and $\delta_c$ are functions of the phase space variables $b$, $c$ and their corresponding canonical conjugates $p_b$ and $p_c$, which  are connected with the spacetime metric through
 Eq.(\ref{metric}). Different choices of $\delta_b$ and $\delta_c$ are frequently referred to as different schemes, and various {ones} are available in the literature. Among them, three major schemes can be identified:
\begin{itemize}
    \item $\mu_o$(-like) schemes: Inspired by the $\mu_o$ scheme in LQC \cite{Ashtekar:2005qt,Ashtekar:2006uz}, in this scheme the polymerization parameters $\delta_b$ and $\delta_c$ are fixed  as constants (see, for example,
\cite{Modesto:2005zm,Modesto:2006mx,Modesto:2008im,Campiglia:2007pr,Corichi:2015xia,
Bodendorfer:2019cyv,Bodendorfer:2019nvy,Assanioussi:2019twp,Gan:2020dkb,ElizagaNavascues:2022npm} and references therein). In these studies, the dependence of $\delta_i$ on some physical parameters involved in the models was considered. However, in all of them, either some inconsistencies or unexpected properties were identified.  In particular, it was found that in some models  physical quantities depend on the choice of the fiducial cell size, while in others singularity resolutions cannot be consistently identiﬁed with a curvature scale. For details, we refer the reader to \cite{Ashtekar:2018cay}. Note that, of the above choices, the Corichi-Singh (CS) model overcomes various problems   \cite{Corichi:2015xia},  but results in a highly asymmetric mass  of the white hole (WH) formed after the bounce. Later, different types of models that result in the symmetric value of masses of BH and WH were developed  \cite{Olmedo:2017lvt,Ashtekar:2018cay}.

    \item $\bar{\mu}$(-like) schemes: Inspired by the $\bar \mu$ scheme in LQC \cite{Ashtekar:2006wn} here the polymerization parameters are chosen to be functions of the phase space variables. This was first implemented for LQBHs in \cite{Boehmer:2007ket} and  adapted later in \cite{Chiou:2008nm}. Although this scheme has alleviated all the shortcomings of the $\mu_o$ scheme in cosmology \cite{Ashtekar:2011ni}, it fails when applied to LQBHs \cite{Ashtekar:2018cay}. In particular, large quantum corrections appear near BH horizons, in which the gravitational fields are already very weak classically. Basically, this quantization does not distinguish between the coordinate and the physical singularity, which is illustrated in Kasner spacetimes too \cite{Singh:2016jsa}. Furthermore, recent work has shown that these effects are, in fact, so strong that  BH/WH horizons do not exist at all \cite{Gan:2022oiy,Zhang:2023noj}. Instead, only transition surfaces exist, and the number of such surfaces are infinite. As a result, the corresponding spacetime is geodesically complete. {This result turns out to be consistent with earlier studies on $\bar \mu$ quantization of Kantowski-Sachs cosmological models with {\emph{matter}} where coordinate singularity does not exist, central singularity is eliminated, and the effective loop quantum spacetime  is geodesically complete  \cite{Saini:2016vgo}.}

    It should be noted that such large quantum effects are closely related to the use of the Kantowski-Sachs metric as describing the internal spacetimes of BHs. In particular, in the Kantowski-Sachs spacetime, the physical length perpendicular to the two-spheres of the classical Schwarzschild BH vanishes not only at the BH singularity but also at its event horizon. Then, the polymerization will lead to significant quantum effects at both locations. To cure the singularity, such large effects  are essential. However, these large effects at BH horizons shall lead  to  inconsistency physically, especially when massive BHs are considered. This is because, near such massive BH  horizons, the gravitational fields are classically very weak, and quantum gravitational effects are expected to be negligible.

    \item Dirac observables: In this scheme, the polymerization parameters are particular functions of the phase space variables such that they are constant along the  trajectories of the dynamical equations.
    {To counter the problems faced by the $\mu_o$ and $\bar{\mu}$ schemes, a Dirac-observable scheme in the extended phase space was first introduced in \cite{Ashtekar:2018lag,Ashtekar:2018cay,Ashtekar:2020ckv}. In these works, the two polymerization parameters were uniquely fixed at the transition surface.} For massive BHs, they are given by \cite{Ashtekar:2018cay}
     \bqn
    \lb{eq1.2}
         \delta^{\text{(AOS)}}_b &=& \left(\frac{\sqrt{\Delta}}{\sqrt{2\pi} \gamma^2 m}\right)^{1/3}, \nb\\
     \delta^{\text{(AOS)}}_c &=& \frac{1}{2L_o}\left(\frac{\gamma\Delta^2}{4\pi^2 m}\right)^{1/3},
    \eqn
where $\gamma$ and $m$  denote, respectively, the Barbero-Immirzi and mass parameters,
$\Delta \left(\equiv 4\sqrt{3} \pi \gamma \ell_{p}^2\right)$ denotes the minimal area gap, $L_o$ is the size of the fiducial cell, and $\ell_{p}$ is the Planck length.
With the above choice of polymerization parameters, it was found that the corresponding LQBH has the following desirable properties: (1) The physical quantities and properties are independent of  $L_o$. (2) The spacetime near the transition surface is symmetric. (3) There exists a universal mass-independent upper bound on the curvature invariants of the effective spacetime at the transition surface.
(4) No mass amplification of the WH exists. (5) There exist negligible quantum corrections at classical scales, including the locations of massive BH/WH horizons.

\end{itemize}

In addition, working within the four-dimensional phase space, spanned by the four Ashtekar phase variables ($b, p_b; c, p_c$), Dirac observables at the Hamiltonian level were also considered in \cite{Bodendorfer:2019xbp,Garcia-Quismondo:2021xdc,Garcia-Quismondo:2022ler,ElizagaNavascues:2022rof,Ongole:2022rqi,ElizagaNavascues:2023gow}.

It is interesting to note that the Ashtekar-Olmedo-Singh (AOS) scheme can be considered as a mixture of the $\mu_o$- and $\bar \mu$ -like schemes. Then, a natural question arises: Are there other choices of polymerization parameters $\delta_b$ and $\delta_c$ for which similar desirable properties as those of the AOS solution exist? The goal of this manuscript will be to explore this avenue in detail. Before we go further, however, it is important to note the following remark.

{\it Remark:} It is to be emphasized that the above comparison between different schemes is restricted to Kruskal spacetimes in GR. Whether the noted advantages and disadvantages extend to other black hole spacetimes and gravitational collapse scenarios is far from clear. In particular, a recent study showed that a $\mu_o$ -like scheme would not permit formation of trapped surfaces in gravitational collapse which a $\bar \mu$ -like scheme permits generically \cite{Li:2021snn}. In addition, the $\bar{\mu}$ scheme was also used in the quantization of the Lemaître-Tolman-Bondi metric  in \cite{Han:2020uhb, Giesel:2021dug, Han:2022rsx}. 
Yet, another different approach was explored in \cite{Gambini:2020qhx,Gambini:2020nsf}, and the model was further studied in \cite{Liu:2021djf}.
To generalize the AOS model, following the line of  previous studies of LQBHs \cite{Gan:2022mle,Gan:2020dkb},
we first make a general survey over the most general solutions of {the effective Hamiltonian which contains five free parameters, including $\delta_b$ and $\delta_c$ which are treated as constant on dynamical trajectories.} Then, using the gauge freedom, we find that one of the five free parameters can be gauged away simply by
the shift of the timelike coordinate, $T \rightarrow T + T_0$. Hence, only four of the five free parameters are physically essential, denoted, respectively, by $c_o$, $m$, $\delta_b$, and $\delta_c$. {It should be noted that  the physical meaning of $c_o$ was studied in detail in   \cite{ElizagaNavascues:2022npm}. However, unlike \cite{ElizagaNavascues:2022npm},}  in our current analysis we do not adopt the AOS choice of  $\delta_b$ and $\delta_c$ given by Eq.(\ref{eq1.2}) and instead consider them as arbitrary constants. Among these four parameters, $m$ is related to the mass parameter of the solution, and $c_o$ characterizes not only the position of the transition surface \cite{ElizagaNavascues:2022npm}, but also the Hawking temperature at the BH horizon. It should be noted that the four-parameter solutions are valid only inside the BH and WH horizons, $T_{\text{WH}} \le T \le T_{\text{BH}}$, where $T_{\text{WH}}$ and $T_{\text{BH}}$ denote the locations of the WH and BH horizon, respectively. To study the asymptotical behavior of the LQBH spacetime, we first extend the solutions analytically beyond these surfaces (cf. Sec.\ref{sec2C}). Since the extension is analytical, it is also unique.

After analytically extending the solutions across the WH and BH horizons,  we find the leading term in the asymptotic expansion of the Kretschmann scalar is $r^{-4}$. This result is well known for the AOS model and we find it to hold for a family of models we study. In addition to the $r^{-4}$ term there also exists an $r^{-5}$ correction along with the $r^{-6}$ term that captures the classical behavior. We find that the $r^{-4}$ term becomes dominant over the $r^{-6}$ term only when $ r > r_{c}$, where $r_{c}$ is the critical radius. In addition, we also calculate the value $r_{c_4}$, the critical radius beyond which the $r^{-4}$ term starts to dominate over the $r^{-5}$ term. Similarly, we calculate $r_{c_5}$ to find the dominant term among $r^{-5}$ and $r^{-6}$ terms. For both the AOS and CS models, we find that $r_c, r_{c_4}, r_{c_5} > L_{\text{obs}}$ for solar mass BHs, where $L_{\text{obs}}$ denotes the size of our observational Universe. This suggests that the asymptotic behavior can still be well described by its classical limit within our observable Universe. In particular, the Kretschmann scalar can be still considered as falling off like $r^{-6}$ in our observable Universe.

On the other hand, requiring that the deviation of the Hawking temperature at the horizon of a massive BH from its classical counterpart be negligible, we find that $c_o$ is uniquely fixed [cf. Eqs. (\ref{eq3.7}) and (\ref{eq3.8})] to
\bqn
\lb{eq1.4}
c_o = \frac{L_o\delta_c \gamma}{8m},
\eqn
which is precisely the choice made in \cite{Corichi:2015xia,Ashtekar:2018lag,Ashtekar:2018cay,Ashtekar:2020ckv}.  Once $c_o$ is fixed, we find that the ratio of the BH and WH masses and the dependence of the curvatures on $m$ at the transition surface are crucially dependent on the choice of the two polymerization parameters $\delta_b$ and $\delta_c$. In particular, we find that as long as they are all inversely proportional to $m^{1/3}$, that is
 \bqn
    \lb{eq1.5}
         \delta_b = \alpha_b \left(\frac{\ell_p}{m}\right)^{1/3}, \;\;\;
         L_o \delta_c = \alpha_c \left(\frac{\ell_p}{m}\right)^{1/3},
    \eqn
where $\alpha_b$ and $\alpha_c$ are two $m$-independent otherwise arbitrary dimensionless  constants,
the  Kretschmann scalar is always independent of $m$ (cf. Fig. \ref{Fig1}), no matter
what the values of $\alpha_b$ and $\alpha_c$ are, although the amplitude of the  Kretschmann scalar indeed depends on specific values of $\alpha_b$ and $\alpha_c$.
On the other hand, when any of the dependence of these two parameters on $m$ is different from $m^{-1/3}$, the amplitude of the  Kretschmann scalar at the transition surface will sensitively depend on the values of $m$,   as shown explicitly in  Figs. \ref{Fig2}-\ref{Fig4}.
 In addition, simply requiring the ratio of the BH and WH masses to be 1 imposes a relation between $\alpha_b$ and $\alpha_c$ [cf. Eq.(\ref{eq3.21})]. In particular,
 assuming that $\delta_b$ takes the form (\ref{eq1.5}), we find that
 $M_{\text{BH}} = M_{\text{WH}}$ leads to
 \bqn
 \lb{eq1.6}
    \alpha_c = \left(\frac{\gamma ^3 \ell_p}{2}\right) \alpha_b^4,\; \left(M_{\text{BH}} = M_{\text{WH}}\right).
\eqn
Therefore, we identify a class of parameters, described by Eqs.(\ref{eq1.4})-(\ref{eq1.6}) with $\alpha_b$ being the free parameter, which share the same properties
 as the AOS solution. The latter corresponds to the particular choice
 \bqn
 \lb{eq1.7}
 \alpha_b^{\text{AOS}} = \left(\frac{\sqrt{\Delta}}{\sqrt{2\pi} \gamma^2 \ell_p}\right)^{1/3}.
 \eqn

The rest of the paper is organized as follows: In Sec.\ref{sec2}, starting with the KS spacetime and its classical Hamiltonian, we first obtain the effective Hamiltonian
by using the replacements (\ref{eq1.1}) and rederive the five-parameter  solutions,
where $B_o, \; p_c^o,\; c_o$ together with $\delta_b$ and $\delta_c$ appearing in Eqs.(\ref{eq2.16}) and (\ref{eq2.17}) are the five parameters. But, as shown there, $B_o$ can be eliminated by the simple replacement $T \rightarrow T + \hat T_o$, so only four of them are physical ones, where $p_c^o$ is related to the mass parameter $m$ via the relation $p_c^o = 4m^2$. In the same section, we consider the marginally trapped surfaces and find that they can happen at $\dot{p}_c = 0$ as well as at $N \rightarrow \infty$.
The former corresponds to the transition surface, and the metric crosses it smoothly, while
the latter represents BH and WH horizons, as shown explicitly in Secs. \ref{sec2B} and \ref{sec2C}.
In Sec. \ref{sec2C}, the metric is also analytically extended beyond these two horizons. In Sec. \ref{sec3A},
spacetimes outside the horizons are studied, including the asymptotic and near horizon regions, while in Sec. \ref{sec3B}, spacetimes across the transition surface and near the WH horizon are investigated. The paper concludes in Sec. \ref{sec4}, where we present our main findings.

\section{loop quantum black holes with constant polymerization parameters} \label{sec2}
\renewcommand{\theequation}{2.\arabic{equation}}
\setcounter{equation}{0}

The spacetime inside a classical spherically symmetric black hole can always be written in the KS form  \cite{Ashtekar:2005qt}
\bq
\lb{metric}
 ds^2 = - N^2 dT^2 + \frac{p_b^2}{L_o^2 \left|p_c\right|} dx^2 + \left|p_c\right| d\Omega^2,
 \eq
 where $N, \; p_b, \; p_c$ are all functions of $T$ only,  and $d\Omega^2 \equiv d\theta^2 + \sin^2\theta d\phi^2$ with $-\infty < T, \; x < \infty$, $\theta \in [0, \pi]$, and $\phi \in [0, 2\pi]$. However, due to the independence of the metric on $x$,  the corresponding Hamiltonian is not well defined, as it is involved in integration over $x$. Then, one usually first introduces a
 fiducial cell with length $L_o$, so that $x \in [0, L_o]$. Physics should not depend on the choice of $L_o$, so at the end of the day we can always take the limit $L_o \rightarrow \infty$, without loss of the generality.
 The functions  $p_b$ and $p_c$ are the dynamical variables, which satisfy the Poisson brackets
 \bqn
 \lb{PBs}
 \{c,p_c \}=2G \gamma,\quad  \{b,p_b \}=G \gamma,
 \eqn
 where $G$ is the Newton gravitational constant, and
  $b$ and $c$ are the corresponding phase space conjugate momenta of $p_b$ and $p_c$, respectively.

It should be noted that the KS metric (\ref{metric}) is invariant under the gauge transformations
 \bq
 \lb{GTs}
 T = f(\hat T), \quad x = \alpha \hat x + x_o,
 \eq
via the redefinition of the lapse function and the length of the fiducial cell,
 \bq
 \lb{RDs}
 \hat N = Nf_{,\hat{T}}, \quad \hat{L}_o = \frac{L_o}{\alpha},
 \eq
 where $f(\hat{T})$ is an arbitrary function of $\hat{T}$,  $\alpha$ and $x_o$ are constants, and $f_{,\hat{T}} \equiv df/d\hat{T}$. Using the above gauge freedom, we can always choose the lapse function as
\bqn
  \lb{eq1}
  N^{\text{GR}}= \frac{\gamma \; {\text{sgn}}(p_c)\left|p_c\right|^{1/2}}{b}.
  \eqn
Then, the corresponding classical Hamiltonian in  GR is given by \cite{Ashtekar:2018cay}
\bqn
  \lb{eq2}
 H^{\text{GR}}[N^{\text{GR}}] &\equiv& N^{\text{GR}} \mathcal{H}^{\text{GR}}\nb\\
&=& -\frac{1}{2G \gamma}\left(2c\;p_c+\left(b+\frac{\gamma^2}{b}\right)p_b\right).~~~
\eqn

\subsection{General spacetimes} \label{sec2A}

To the leading order, it is expected that, as in LQC \cite{Ashtekar:2011ni}, the quantum effects can be well captured
by replacing the two canonical phase space variables $b$ and $c$ via the relations given by Eq.(\ref{eq1.1})
in the classical lapse function and  Hamiltonian \cite{Ashtekar:2004eh,Ashtekar:2011ni}, where
 the two polymerization parameters $\delta_b$ and $\delta_c$ are,  in general, functions of the phase space variables ($b, p_b; c, p_c$). However, in the current paper we shall focus ourselves only on the cases in which they are constants.
Inserting the above replacements into Eqs.(\ref{eq1}) and (\ref{eq2}) we find that
the effective Hamiltonian for LQBHs is given by
\bqn
\label{Heff}
H_{\mathrm{eff}}  &\equiv& N \mathcal{H} = \frac{L_{o}}{G}\left(O_{b}-O_{c}\right),
\eqn
with
\bqn
\label{lapse}
N &=& \frac{\gamma \delta_b\; {\text{sgn}}\left(p_c\right) \sqrt{|p_c|}}{\sin{\left(\delta_b b\right)}}, \nb\\
 O_{b} &\equiv&-\frac{p_{b}}{2 \gamma L_{o}}\left(\frac{\sin \left( \delta_{b} b\right)}{\delta_{b}}+\frac{\gamma^{2} \delta_{b}}{\sin \left(\delta_{b} b\right)}\right),\nb\\
    O_{c} &\equiv& \frac{\left|p_{c}\right|}{\gamma L_{o}} \frac{\sin\left( \delta_{c} c\right)}{\delta_{c}}.
\eqn
It can be shown that $O_{b}$ and $O_{c}$ are the only two independent Dirac observables that can be constructed in the current case.
Note that $\sqrt{|p_c|}$ is the geometric radius of the two-spheres with $T, x = $ constant. Thus, without loss of the generality (as far as the effective semiclassical approximations are concerned), we can always assume that $p_c \ge 0$. Then, the corresponding Hamiltonian equations for the four phase variables
 $(b, p_b; c, p_c)$ are given, respectively, by
\bqn
\lb{eq2.14}
\dot{c} &=& - 2 \frac{\sin\left(\delta_c c\right)}{\delta_c}, \\
\lb{eq2.15}
\dot{p}_c &=&{2}p_c\cos\left(\delta_c c\right),\\
\lb{eq2.12}
\dot{b} &=& - \frac{1}{2}\left(\frac{\sin\left(\delta_b b\right)}{\delta_b} + \frac{\gamma^2\delta_b}{\sin\left(\delta_b b\right)}\right), \\
\lb{eq2.13}
\dot{p}_b &=& \frac{1}{2}p_b\cos\left(\delta_b b\right)\left(1  -  \frac{\gamma^2\delta_b^2}{\sin^2\left(\delta_b b\right)}\right).
\eqn
The integration of the first three equations yields \cite{ElizagaNavascues:2022npm}
\bqn
\lb{eq2.16}
     \sin\left(\delta_c c\right) &=& \frac{2 c_o e^{-2T}}{1 + c_o^2 e^{-4T}},\nb\\
     p_c &=& p_{c}^{o} \left(c_o^2 e^{-2T} + e^{2T}\right), \nb\\
     \cos\left(\delta_b b\right) &=& b_o \tanh\left(\frac{b_{o} T}{2} + B_o\right),
\eqn
where $c_o,\; p_c^o,\; B_o$ are integration constants, and

\bqn
\lb{eq2.16a}
b_o \equiv \sqrt{1+\gamma^2 \delta_b^2}.
\eqn
Substituting the above solutions to the effective Hamiltonian (\ref{Heff}), we find that
\bqn
\lb{eq2.17}
    p_{b} &=& - p_b^o \cosh^2 \left(\frac{b_{o} T}{2} + B_o\right)\nb\\
    &&~~~~ \times \Bigg[1- b_o^2 \tanh^2\left(\frac{b_{o} T}{2} + B_o\right)\Bigg]^{1/2}, ~~~
\eqn
where $p_b^o \equiv \frac{4 c_o p_{c}^{o} \delta_b }{b_o^2 \delta_{c}}$. However, since we assume $p_c \ge 0$, we must require $p_c^o > 0$.
Corresponding to the above solution, it can be shown that the two Dirac observables are given by
\bqn
\label{Ob}
    O_{b} = O_{c} = \frac{2 c_o p_{c}^{o}}{\gamma L_o  \delta_{c}}.
\eqn
It can be also shown that
\bqn\lb{MCs}
    N^2 &=&  p_{c}^{o} \gamma^{2} \delta_{b}^2 \frac{c_o^2 e^{-2T} + e^{2T}}{1- b_o^2 \tanh^2\left(\frac{b_{o} T}{2} + B_o\right)}, \nb\\
    g_{xx} &=&  \alpha^2  \cosh^4 \left(\frac{b_{o} T}{2} + B_o\right)\nb\\
    && \times \frac{1- b_o^2 \tanh^2\left(\frac{b_{o} T}{2} + B_o\right)}{c_o^2 e^{-2T} + e^{2T}},
\eqn
where $\alpha^2 \equiv  \frac{16  p_{c}^{o} c_o^2  \delta_{b}^2 }{b_o^4 L_o^2  \delta_{c}^2 }$.
It is clear that the above solutions contain five free parameters ($c_o,p_c^o,B_o,\delta_b,\delta_c$). By using the following arguments we show that only four are physical.
\begin{itemize}
    \item From Eq.(\ref{eq2.16}) we can see that $p_c$ is always positive and nonzero. In fact, at $T_{\cal{T}} \equiv (1/4)\ln c_o^2$  it reaches its minimal value
    \bqn
    \lb{p3219}
    p_c^{\cal{T}}\left(T_{\cal{T}}\right) \equiv 2p_c^o |c_o|.
    \eqn
    Apart from it, $p_c$ is increasing in both directions, $T > T_{\cal{T}}$ and $T < T_{\cal{T}}$. Thus, the surface $T = T_{\cal{T}}$ acts as a transition surface, which will be denoted as surface ${\cal{T}}$.

    \item The lapse function becomes unbounded at
    \bqn
    \lb{eq2.18}
    T_{\pm} = \frac{1}{b_o}\left(\ln\frac{b_o \pm 1}{b_o \mp 1} - 2B_o\right).
    \eqn
    Clearly, these two singularities restrict the above solutions to the region
    $ T_{-} < T < T_+$. Then, extensions are needed beyond these two surfaces in order to obtain a geodesically complete spacetime, provided that the spacetime is not singular at these two points. As a matter of fact, these surfaces represent the black and white hole horizons, respectively \cite{ElizagaNavascues:2022npm}, and the extensions can be easily carried out as in the classical case. In particular, requiring the extension be analytical, it is also unique, as in the classical case.

\end{itemize}

Before showing our above claims, let us first simplify the above solutions by using the gauge residuals left from the gauge freedom (\ref{GTs}). First, by the shift symmetry $T \rightarrow \hat{T} = T + \hat{T}_o$, we find that the metric (\ref{metric}) takes the form
$ds^2 = - \hat{N}^2 d\hat{T}^2 + \hat{g}_{xx} dx^2 + \hat{p}_cd\Omega^2$,
where $\left(\hat{N}^2, \; \hat{g}_{xx}, \; \hat{p}_c\right)$ are given by Eqs.(\ref{eq2.16}) and
(\ref{MCs}) with the replacements
\bqn
\lb{Replacements}
\left(T, B_o, p_c^o, c_o, L_o, \alpha\right) \rightarrow
\left(\hat T, \hat B_o, \hat p_c^o, \hat c_o, \hat L_o, \hat \alpha\right),
\eqn
where
\bqn
\lb{eq2.21}
\hat B_o  &=& B_o - \frac{b_o}{2}\hat T_o, \quad
\hat p_c^o = e^{-2\hat T_o} p_c^o, \nb\\
\hat c_o  &=& c_o e^{2\hat T_o}, \quad
\hat L_o =  L_o, \quad \hat{\alpha} = \alpha e^{\hat T_o}.
\eqn
Since  $\hat T_o$ is an arbitrary constant, without loss of the generality, we can always set it to
\bqn
\lb{eq2.19}
\hat{T}_o = \frac{2}{b_o}\left(B_o - \tanh^{-1}\left(\frac{1}{b_o}\right)\right).
\eqn
Then, we find that in terms of $\hat{T}$, the two horizons given by Eq.(\ref{eq2.18}) now are located, respectively, at
\bqn
\lb{eq2.20}
\hat T_{\text{BH}} \equiv \hat T_+ &=& 0, \nb\\
\hat T_{\text{WH}} \equiv \hat T_- &=& - \frac{2}{b_o}\ln\frac{b_o +1}{b_0 -1} < 0.
\eqn
As indicated by their subscriptions, $\hat T_{\text{BH}}$ and $\hat T_{\text{WH}}$  will correspond to the locations of the black and white hole horizons, respectively.
The above choice is also consistent with that adopted in \cite{Corichi:2015xia,Ashtekar:2018cay}, so that the coordinate $\hat T$ is strictly negative between the black and white hole horizons.  It is interesting to note that
$T_{\text{WH}} \rightarrow -\infty$ as $b_o \rightarrow 1$ (or $\delta_b \rightarrow 0$),
which corresponds to the classical limit, and  the WH horizon turns into the spacetime singularity, at which now we have $\left.p_c(T = -\infty)\right|_{\delta_b = 0} = 0$.

On the other hand, considering the rescaling freedom of Eq.(\ref{GTs}) for the $x$ coordinate,  the metric  can be finally cast in the form
\bqn
\lb{metricB}
ds^2 = - \hat{N}^2 d\hat{T}^2 + \hat{g}_{xx} d\hat x^2 + \hat{p}_cd\Omega^2,
\eqn
with
\bqn\lb{MCsA}
    \hat N^2 &=&  \hat p_{c}^{o} \gamma^{2} \delta_{b}^2 \frac{\hat c_o^2 e^{-2\hat T} + e^{2 \hat T}}{1- b_o^2 \tanh^2\left(\frac{b_{o} \hat T}{2} + \hat B_o\right)}, \nb\\
    \hat g_{xx} &=&  \gamma^2\delta_b^2 \cosh^4 \left(\frac{b_{o} \hat T}{2} + \hat B_o\right)\nb\\
    && \times \frac{1- b_o^2 \tanh^2\left(\frac{b_{o} \hat T}{2} + \hat B_o\right)}{\hat c_o^2 e^{-2\hat T} + e^{2\hat T}},\nb\\
 \hat p_c &=& \hat p_{c}^{o} \left(\hat c_o^2 e^{-2\hat T} + e^{2\hat T}\right),\nb\\
 \hat B_o &=&  \tanh^{-1}\left(\frac{1}{b_o}\right).
\eqn
Note that in the expression of $\hat g_{xx}$ the factor $\gamma^2\delta_b$ is kept, as this will allow us to take the classical limit $\delta_b \rightarrow 0$, considering the fact
\bqn
\lb{eq2.25aa}
&& \cosh^2\left(y + \hat B_o\right) = \frac{\cosh^2 y}{\gamma^2\delta_b^2}\left(b_o + \tanh y\right)^2,\nb\\
&& 
1- b_o^2 \tanh^2\left(y + \hat B_o\right) = -
\frac{\gamma^2\delta_b^2}{\left(b_o + \tanh y\right)^2} \nb\\
&& ~~~~~~~~~~~~~\times \tanh y \left[2b_o + \left(b_o^2 + 1\right) \tanh y\right].~~~~~~
\eqn
Therefore, from Eq. (\ref{eq2.16a}) we find that
in the current case {\em there are only four essential parameters, $\hat{p}_c^o, \; \delta_b, \; \delta_c$, and $\hat{c}_o$, which uniquely determine the properties of the spacetimes}. In addition, if we require that the above solutions will reduce to the classical one as $\delta_b, \; \delta_c \rightarrow 0$, we must require $\hat{c}_o = \hat{c}_o\left(\delta_c\right)$ and
\bqn
\lb{eq2.22}
\lim_{\delta_c \rightarrow 0} \hat c_o\left(\delta_c\right) = 0,
\eqn
as can be seen from Eqs. (\ref{eq2.14}), (\ref{eq2.15}), and (\ref{eq2.16}).
Moreover, in the classical limit the BH horizon is located at $\hat{p}^{\text{GR}}_c(0) = (2m)^2$, which corresponds to $\hat{p}_c^o = 4 m^2$ and shall be adopted in the rest of this paper.

It is interesting to note that in \cite{Corichi:2015xia,Ashtekar:2018cay}
$\hat{c}_o$ was chosen as
\bqn
\lb{eq2.23}
\hat{c}_o^{\text{(CS, AOS)}} = \frac{\gamma L_o\delta_c}{8 m},
\eqn
which clearly satisfies Eq.(\ref{eq2.22}). With this choice, Eq.(\ref{Ob}) yields
\bqn
\label{Ob_CS}
    \hat O^{\text{(CS, AOS)}}_{b} = \hat O^{\text{(CS, AOS)}}_{c} = m.
\eqn
However, in the following we shall leave this possibility open, and only impose the condition (\ref{eq2.22}). {Moreover,
for massive BHs, in \cite{Ashtekar:2018cay} $\delta_b$ and $\delta_c$ were chosen as that given in Eq.(\ref{eq1.2}).}
In addition, in \cite{Corichi:2015xia} the following choice was adopted
\bqn
\label{Deltas_CS}
    \delta^{\text{(CS)}}_b = \frac{\sqrt{\Delta}}{r_o}, \quad
     \delta^{\text{(CS)}}_c = \frac{\sqrt{\Delta}}{L_o},
\eqn
where $r_o$ denotes the geometric radius of the fiducial metric $ds_o^2 = d x^2 + r_o^2d^2\Omega$ considered in \cite{Corichi:2015xia}. Again, in this paper we shall also leave these choices open, in order to have a general survey of the four-dimensional phase space.

In addition, without causing any confusion, in the rest of this paper we shall drop all superscript hats in Eqs. (\ref{metricB})-(\ref{eq2.22}), so the spacetimes to be considered in the rest of this paper are described by
the metric
\bqn
\lb{metricC}
ds^2 = - {N}^2 d{T}^2 + {g}_{xx} dx^2 + {p}_cd\Omega^2,
\eqn
with
\bqn
\lb{eq2.31}
 N^2 &=& \frac{p_c{\cal{D}}^2}{\cal{A}}, \nb\\
 g_{xx} &=& \frac{4m^2 \left[b_o\cosh\left(\frac{b_o T}{2}\right)
 + \sinh\left(\frac{b_o T}{2}\right)\right]^4}{p_c{\cal{D}}^2}{\cal{A}},\nb\\
 p_c &=&  4m^2 \left(c_o^2 e^{-2T} + e^{2T}\right),
 \eqn
where
\bqn
\lb{eq2.30}
 {\cal{A}} &\equiv& 2\left(b_o^2 + 1\right) - \left(b_o+1\right)^2 e^{b_o T} - \left(b_o-1\right)^2 e^{-b_o T},\nb\\
 {\cal{D}} &\equiv& \left(b_o+1\right) e^{b_o T/2} + \left(b_o-1\right) e^{-b_o T/2}.
 \eqn

Now let us turn to our previous  claims regarding the existence of transition surface and BH and WH horizons. To this goal, let us first consider the existence of a ``marginally trapped surface." The latter can be found by calculating  the expansions of the ingoing and outgoing radially moving light rays \cite{Baumgarte:2010ndz,Hawking:1973uf,Ashtekar:2018cay,Wang:2003bt,Wang:2003xa}. Introducing the unit vectors, $u_{\mu} \equiv N \delta^T_{\mu}$ and $s_{\mu} \equiv \sqrt{g_{xx}}\delta^x_{\mu}$, we  construct  two null vectors $\ell_{\mu}^{\pm} = \left(u_{\mu} \pm s_{\mu} \right)/\sqrt{2}$, which define, respectively, the ingoing and outgoing radially moving light rays. Then, the expansions of these light rays are given by
\bq
\lb{eq2.24}
\Theta_{\pm} \equiv m^{\mu\nu}\nabla_{\mu}\ell_{\nu}^{\pm} = - \frac{\dot p_{c}}{\sqrt{2} N p_c},
\eq
 where $m_{\mu\nu} \equiv g_{\mu\nu} + u_{\mu}u_{\nu} - s_{\mu}s_{\nu}$.
{A marginally trapped surface is defined as the location at which    $\Theta_{+} \Theta_{-} = 0$ \cite{Hawking:1973uf,Ashtekar:2018cay,Wang:2003bt,Wang:2003xa,Gong:2007md}.
Clearly, in the current case this is possible only when (a) $\dot p_{c} = 0$ or (b) $N = \infty$. In the following, let us consider them separately.

\subsection{Transition surface}
\label{sec2B}

From Eq.(\ref{MCsA}) we find that
\bqn\lb{eq2.25}
\dot p_{c} &=&  \frac{8m^2}{e^{2T}}\left( e^{4T} - c_o^2\right)
=\begin{cases}
> 0, & T > T_{\cal{T}},\cr
= 0 , & T = T_{\cal{T}},\cr
< 0, & T < T_{\cal{T}},\cr
\end{cases}
~~~~~
\eqn
where  
\bqn
\lb{eq2.25b}
T_{\cal{T}} = \frac{1}{2}\ln c_o.
\eqn
 It is clear that in the region $T > T_{\cal{T}}$ both of $\Theta_{\pm}$ are negative, so the spacetime in this region is ``trapped." On the other hand, in the region
  $T < T_{\cal{T}}$ both of them are positive, and the corresponding spacetime  now becomes ``antitrapped." In addition, the metric coefficients are {smooth} across $T = T_{\cal{T}}$.  Therefore, this marginally trapped surface is a transition surface that separates a trapped region from an antitrapped one. The geometric radius
  $\sqrt{p_c}$ of the two-spheres is increasing apart from this surface in both directions. At this transition surface, the area of the two-spheres   reaches its minimal value,
 \bqn
\lb{eq2.26}
A^{\text{min}} \equiv  4\pi  p_c\left(T_{\cal{T}}\right) = 32\pi m^2 c_o.
\eqn
For the choice of Eq.(\ref{eq2.23}), we have
 \bqn
\lb{eq2.27}
A^{\text{min}}_{\text{(AOS)}}  &=&    4\pi m \gamma L_o \delta_c\nb\\
&=&  4  \pi \gamma^2 \left(\frac{3}{2}\right)^{1/3}     \left(\frac{m}{\ell_{pl}}\right)^{2/3} \; \ell_{pl}^2.
\eqn
Note that in the last step of the above equation, we used the value of $L_o\delta_c$
given by Eq.(\ref{eq1.2}) for massive black holes \cite{Ashtekar:2018cay}. Thus, for solar {mass} black holes, we have
\bqn
\lb{eq2.28}
\left. A^{\text{min}}_{\text{(AOS)}}\right|_{m \gtrsim
 m_{\bigodot}} \gtrsim 10^{25} \ell_{pl}^2.
\eqn

 \subsection{Black and white hole horizons and analytical extensions beyond them}
 \label{sec2C}

As noted above, the lapse function $N$ becomes unbounded at $T = T_{\text{BH}}$
and $T = T_{\text{WH}}$, where $T_{\text{BH}}$ and $T_{\text{WH}}$ are given by Eq.(\ref{eq2.20}). Clearly, on these surfaces, $\Theta_{\pm} = 0$. So, they also represent marginally trapped surfaces. To see the nature of these surfaces, let us first note that ${\cal{D}}$ defined above is strictly positive ${\cal{D}} > 0$, while ${\cal{A}}$ vanish at the two points defined by Eq.(\ref{eq2.20}).
 In fact, ${\cal{A}}(T)$ can be written as
 \bqn
 \lb{eq2.32}
 {\cal{A}}(T) =  \frac{\left(b_o+1\right)^2}{e^{b_o T}}\left(1 - e^{b_o T}\right) \left(e^{b_o T} - e^{b_o T_{\text{WH}}}\right).
 \eqn

 It can be shown that the singularities at $T = 0$ and $T =  T_{\text{WH}}$ are coordinate ones, similar to the classical Schwarzschild solution at $r = 2m$. In particular, near these singularities we have $N^2 \propto \left(e^{b_o T} - e^{b_o T_{\text{A}}}\right)^{-1}$ and
 $g_{xx} \propto \left(e^{b_o T} - e^{b_o T_{\text{A}}}\right)$,
 where $T_{\text{A}}$ denotes one of the locations of the two horizons. To make extensions across each of them, it is sufficient to consider the neighborhood of these horizons, at which we find that
 \bqn
 \lb{eq2.33}
 N^2 &\simeq& \frac{b_o^2 f_1(T_A)e^{2b_o T}}{e^{b_o T} - e^{b_o T_{\text{A}}}}, \nb\\
 g_{xx} &\simeq& f_2(T_A) \left(e^{b_o T} - e^{b_o T_{\text{A}}}\right),
 \eqn
 where 

\begin{align}
\lb{eq2.33aa}
     f_1(T_{BH})&=4 b_o \left(c_o^2+1\right) m^2 , \nb\\
    f_1(T_{WH})&=4 m^2 b_o \left(\frac{b_o-1}{b_o+1}\right)^{\frac{4}{b_o}-2} \nb\\
    & ~~~ \times \left(\left(\frac{b_o+1}{b_o-1}\right)^{\frac{8}{b_o}} c_o^2+1\right), \nb\\
    f_2(T_{BH})&=\frac{b_o^3}{\left(c_o^2+1\right)}, \nb\\
    f_2(T_{WH})&=\frac{b_o^3 \left(\frac{b_o+1}{b_o-1}\right)^{\frac{4}{b_o}+2}}{\left(\frac{b_o+1}{b_o-1}\right)^{\frac{8}{b_o}} c_o^2+1}.
\end{align}
  It is interesting to note that,  when $b_o = 1$, $f_{1,2}(T_{\text{BH}})$ reduce to those of the classical BH, while $f_{1,2}(T_{\text{WH}})$ diverge under this limit. This is because now the WH horizon turns into the classical spacetime singularity, as noted previously.

Then, we find that the 2D metric in the ($T, x)$ plane takes the form
  \bqn
 \lb{eq2.34}
 d\Sigma_2 \simeq f_2(T_A)\left(e^{b_o T} - e^{b_o T_{\text{A}}}\right)\left(-dt^2 + dx^2\right),
 \eqn
where
\bqn
\lb{eq2.35}
 t \simeq f(T_A)\ln\left(e^{b_o T} - e^{b_o T_{\text{A}}}\right) + t_A,
 \eqn
 with $f(T_A)  \equiv \sqrt{f_1/f_2}$, and $t_A$ being a constant. Setting $v = t + x, w = t - x$, we find that
\bqn
 \lb{eq2.36}
 d\Sigma_2 \simeq -  f_2(T_A)\left(e^{b_o T} - e^{b_o T_{\text{A}}}\right)dv dw.
 \eqn

 To eliminate the coordinate singularity, we further introduce $\hat v$ and $\hat w$ via the relations $\hat v = e^{a v}, \; \hat w = e^{a w}$,
 so that
 \bqn
 \lb{eq2.37}
 d\Sigma_2 \simeq - \frac{ f_2(T_A)}{a^2e^{2a t_A}}\left(e^{b_o T} - e^{b_o T_{\text{A}}}\right)^{1- 2a f(T_A)} d\hat v d\hat w. ~~~
 \eqn
Clearly, choosing
 \bqn
 \lb{eq2.38}
 a = \frac{1}{2 f(T_A)},
 \eqn
we find that the coordinate singularity disappears in the ($\hat v, \hat w$) coordinates. On the other hand, since the functions
${\cal{D}}, \; {\cal{A}}, \; p_c$ and $\cosh(b_oT/2 + B_o)$ are all analytical functions of $T$, it is not hard to be convinced that such extensions are analytical across each of these two singular surfaces.  As a result, these extensions are also unique, and the extended spacetimes take precisely the form of
Eqs.(\ref{metricB}) and (\ref{MCsA}) in the ($T, x, \theta, \phi$) coordinates in each of the three regions: $-\infty < T < T_{\text{WH}}$, $ T_{\text{WH}} < T < T_{\text{BH}}$ and
$T_{\text{BH}} < T < \infty$. Across the two horizons,
the spacetimes are smoothly connected by the ($\hat v, \hat w$) coordinates.

\section{Local and Global Properties of the extended spacetimes} \label{sec3}
\renewcommand{\theequation}{3.\arabic{equation}}
\setcounter{equation}{0}

In this section, we shall study the local and global properties of the extended spacetimes in detail. The extensions introduced in the last section  allow us to study the spacetimes  in each of the three regions separately.

\subsection{Spacetime outside of the BH horizon} \label{sec3A}

Because of the choice of the free parameter $\hat T_o$ of Eq.(\ref{eq2.19}), the black hole horizon now is located at $T_{\text{BH}} = 0$. Then, from Eq.(\ref{eq2.32}) we find that
\bqn
 \lb{eq3.1}
 {\cal{A}}_{+}(T) &\equiv& -  {\cal{A}}(T) = \frac{\left(b_o+1\right)^2}{e^{b_o T}}\nb\\
 && \times \left(e^{b_o T} - 1\right) \left(e^{b_o T} - e^{b_o T_{\text{WH}}}\right)  > 0, ~~~
 \eqn
 for $T > 0$. Thus, the normal vector $N^{+}_{\mu} \equiv \delta^T_{\mu}$ to the hypersurface $T =$ constant becomes spacelike, as now we have
 \bqn
 \lb{eq3.2}
 g^{\mu\nu}N^{+}_{\mu} N^{+}_{\nu} =  \frac{{\cal{A}}_+}{p_c {\cal{D}}^2} > 0.
 \eqn
 As a result, the metric in this region takes the form
 \bqn
\lb{eq3.3}
ds^2_{+} = - N^2_{+} dx^2 +  \frac{p_c {\cal{D}}^2}{{\cal{A}}_{+}}  dT^2 + {p}_c d\Omega^2,
\eqn
with
\bqn\lb{eq3.4}
    N_{+}  &\equiv&   \frac{2m \sqrt{{\cal{A}}_+}}{\sqrt{p_c} {\cal{D}}} \cosh^2\left(\frac{b_o T}{2}\right)\nb\\
    && \times \left[b_o + \tanh\left(\frac{b_oT}{2}\right)\right]^2. ~~~~~
\eqn

Introducing the two unit vectors, $u^{+}_{\mu} \equiv N_{+}\delta_{\mu}^{x}$ and
$s^{+}_{\mu} \equiv \sqrt{g_{TT}}\delta_{\mu}^T$, we can construct two null vectors,
$\ell_{\mu}^{(+, \pm)} \equiv \left(u^{+}_{\mu} \pm s^{+}_{\mu}\right)/\sqrt{2}$, which define, respectively, the ingoing and outgoing radially moving light rays. Then, the expansions of these light rays are given by
\bq
\lb{eq3.5}
\Theta^{(+)}_{\pm} \equiv m^{\mu\nu}\nabla_{\mu}\ell_{\nu}^{(+, \pm)} = \pm
\sqrt{\frac{{\cal{A}}_+}{2p_c {\cal{D}}^2}}\; \frac{\dot p_{c}}{p_c},
\eq
 with
 \bq
\lb{eq3.6}
\dot p_{c} = 8m^2 e^{-2T}\left(e^{4T} - c_o^2\right) > 0,
\eq
for $T> 0$, provided that $c_o^2 < 1$.  This is always the case, as long as the temperature of massive BHs does not deviate significantly from its classical value, as to be shown below. Therefore, in this region the expansion of the ingoing null rays is always negative, while the expansion of the outgoing null rays is always positive, so the spacetimes in this region are untrapped \cite{Hawking:1973uf,Wang:2003bt,Wang:2003xa}. Thus, the horizon located at $T = 0$ separates the untrapped region ($T > 0$) from the trapped one ($T_{\text{WH}} < T < 0$), and acts like a BH horizon, although in the trapped region the spacetime is free of any kind of spacetime singularities. The analytic extension can be carried out by introducing the ($\hat v, \hat w$) coordinates, presented in the last section.

 To study further the properties of the spacetimes in this region, let us consider  the asymptotic behaviors ($T \rightarrow \infty$) of the spacetimes and their near horizon properties ($T \gtrsim 0$), separately. {We majorly use the Kretschmann scalar for analyzing the behavior of the spacetime and this has been calculated using the xAct package in {\em Mathematica.}}

 \subsubsection{Asymptotic behavior of the spacetime} \label{sec3A2}

As $T \rightarrow \infty$, we find that the corresponding Kretschmann scalar is given by
\begin{widetext}
\begin{align}
     \label{eq3.9}
     K \equiv R_{\alpha\beta\delta\gamma}R^{\alpha\beta\delta\gamma} =  \frac{a_0}{r^4} + \frac{a_1}{r^{4 + b_o}} + \frac{a_2}{r^{4+2b_o}} + \frac{a_3}{r^{4+3b_o}} + \frac{a_4}{r^{8}} + \mathcal{O} \left( \frac{1}{r^{4(1+b_o)}}\right),
 \end{align}
 where $r \equiv 2m e^T$ and
 \begin{align}
 \lb{eq3.9aa}
     a_0 &\equiv 4\left(b_o-1\right)^2 \left(b_o^2-4 b_o+6\right), \nb\\
     a_1 &\equiv -\frac{16(2m)^{b_o} b_o \left(b_o-1\right) \left(b_o^4-7 b_o^3+17 b_o^2-7 b_o-8\right)}{\left(b_o+1\right)^2}, \nb\\
     a_2 &\equiv \frac{16(2m)^{2b_o} b_o \left(b_o^7-21 b_o^6+90 b_o^5-84 b_o^4-57 b_o^3+89 b_o^2+2 b_o-8\right)}{\left(b_o+1\right)^4}, \nb\\
     a_3 &\equiv \frac{16(2m)^{3b_o}\left(b_o-1\right) b_o \left(5 b_o^7+28 b_o^6-265 b_o^5+284 b_o^4+103 b_o^3-184 b_o^2-3 b_o+8\right)}{\left(b_o+1\right)^5}, \nb\\
     a_4 &\equiv -128 m^4\left(b_o-1\right) \left(b_o^3-11 b_o^2+42 b_o-46\right) c_o^2.
 \end{align}
\end{widetext}
It is interesting to note that the above expressions are valid for any choice of $(\delta_{b}, \delta_{c},  c_o, m)$. In particular, they reduce precisely to its classical limit, $K_{\text{GR}} = 48m^2/r^6$, when $\delta_b = 0$, independent of the choices  of
$\delta_{c}, c_o$ and $m$. In fact, when $\delta_b = 0$ we have $b_o = 1$, and then from the above expressions we find that $a_2 = 48m^2$, while all other terms identically vanish.

 On the other hand, $a_0 \not= 0$, as long as $\delta_b \not=0$. That is, the Kretschmann scalar is asymptotically vanishing as $r^{-4}$, instead of $r^{-6}$, as that in the classical case. This is true not only in the $\mu_0$ scheme, but also true in the general case, in which $\delta_b$ and $\delta_c$ are the Dirac observables of the four-dimensional phase spacetime ($b, p_b; c, p_c$) \cite{Ongole:2022rqi}.
{

However, it must be noted that  the $r^{-4}$ term becomes dominant only when
$r > r_c$, where $r_c$ is defined by
\bqn
\lb{eq3.10a}
\frac{a_0}{r^4_c} = \frac{a_2}{r^{4+2b_o}_c}.
\eqn
In particular, for the AOS choice in Eq.(\ref{eq1.2}), we find that in the large mass limit
 \bqn
 \lb{eq3.10b}
    r_c^{AOS} &\simeq& \left(\frac{8\left(2/3\right)^{1/3}}{\gamma ^2}\right)^{1/2} \frac{m^{5/3}}{\ell_p^{2/3}}  + \mathcal{O}\left(m^{4/3}\right). ~~~~
\eqn
Thus, for solar mass BHs, we have
 \bqn
 \lb{eq3.10c}
    r_{c, M_{\odot}}^{AOS} &\simeq&  2.06 \times 10^{31} \; {\text{cm}} > L_{\text{obs}},  ~~~~
\eqn
where $L_{\text{obs}}\; (\simeq 4.4 \times 10^{28} \; {\text{cm}})$ denotes the  size of our observational Universe.

Similarly, we can define the critical value $r_{c_4}$, beyond which the $r^{-4}$ starts to dominate over the $r^{-5}$ term, and $r_{c_5}$ as the critical radius beyond which the $r^{-5}$ term becomes dominant over the $r^{-6}$ term. These are defined by
\begin{align}
    \label{eq3.15}
    \frac{a_0}{r^{4}_{c_4}} &= \frac{a_1}{r^{4+b_o}_{c_4}} \nb\\
    \frac{a_1}{r^{4+b_o}_{c_5}} &= \frac{a_2}{r^{4+2b_o}_{c_5}}
\end{align}

These expressions in the large mass limit for the AOS choice in Eq.(\ref{eq1.2}) are as follows
\begin{align}
    \label{eq3.15a}
     r_{c_4}^{AOS} &\simeq \frac{2^{11/3} \; m^{5/3}}{3^{7/6} \; \gamma \; \ell_p^{2/3}} + \mathcal{O}\left(m\right), \nb\\
     r_{c_5}^{AOS} &\simeq \frac{3^{5/6} m^{5/3}}{{2}^{1/3} \gamma {\ell_p^{2/3}}} + \mathcal{O}\left(m\right).
\end{align}
And for a solar mass BH, we have
\begin{align}
    \label{eq3.15b}
     r_{c_4, M_{\odot}}^{AOS} &\simeq 2.75 \times 10^{31} \; {\text{cm}} > L_{\text{obs}}, \nb\\
     r_{c_5, M_{\odot}}^{AOS} &\simeq 1.54 \times 10^{31} \; {\text{cm}} > L_{\text{obs}} .
\end{align}

It is clear from the above calculations that $r^{-4}$ and $r^{-5}$ terms will be dominant at distances much beyond our observational Universe. Hence, for BHs with solar mass and greater, the asymptotic behavior of the AOS spacetime can be well described by its classical limit ($K \simeq r^{-6}$) within our observational Universe.}

{ On the other hand, for the CS choice (\ref{Deltas_CS}), we find
\bqn
\lb{eq3.10g}
    r_c^{CS} &\simeq& \left(\frac{16}{3 \pi ^2 \gamma ^6 }\right)^{1/2}\frac{m^3}{\ell_p^2}  + \mathcal{O}\left(m^{2}\right),  \nb\\
    r_{c_4}^{CS} &\simeq&   \frac{2^4 m^3}{3^{3/2} \pi  \gamma ^3 \ell_p^2} + \mathcal{O}\left(m\right), \nb\\
    r_{c_5}^{CS} &\simeq& \frac{\sqrt{3} m^3}{\pi  \gamma ^3 \ell_p^2}  + \mathcal{O}\left(m\right), 
\eqn
and for a solar mass BH we obtain $r_{c, M_{\odot}}^{CS} \simeq  4.18 \times 10^{82} \; {\text{cm}}$, $r_{c_4, M_{\odot}}^{CS} \simeq  5.57 \times 10^{82} \; {\text{cm}} $ and $r_{c_5, M_{\odot}}^{CS} \simeq  3.13 \times 10^{82} \; {\text{cm}} \gg L_{\text{obs}}$.}
{
Therefore, in both models  the asymptotic behavior of the spacetime can be well described by its classical limit ($K \simeq r^{-6}$) within our observational Universe.}

To understand further the asymptotic behavior of the spacetimes, let us calculate  the effective energy-momentum tensor,  given by

\begin{align}
    \label{eq3.16}
    \kappa T_{\mu\nu} \equiv G_{\mu\nu} =  \rho u_{\mu}u_{\nu}  + p_r r_{\mu}r_{\nu} + p_{\bot}\left(\theta_{\mu}\theta_{\nu} + \phi_{\mu}\phi_{\nu}\right),
\end{align}
{where $u_{\mu}$ denotes the unit timelike vector along the $T$ direction, and $r_{\mu}$,   $\theta_{\mu}$, and $\phi_{\mu}$ are the spacelike unit vectors along $r$, $\theta$, and $\phi$ directions respectively. In addition $\rho$, $p_r$, and $p_{\bot}$ are the energy density and pressures along the radial and tangential directions and $\kappa =8\pi G$.} To the leading order, they are given by

\begin{align}
\label{eq3.17}
    \rho &\simeq -\frac{4(2m)^{b_o}\left(b_o-1\right) b_o^2}{r^{2+b_o} \left(b_o+1\right)^2} + \mathcal{O} \left( \frac{1}{r^{2(1+b_o)} } \right), \\
    p_r &\simeq  \frac{2(b_o-1)}{ r^2} + \mathcal{O} \left( \frac{1}{r^{2+b_o}} \right), \\
    p_{\bot} &\simeq \frac{\left(b_o-1\right)^2}{r^2} + \mathcal{O} \left( \frac{1}{r^{2+b_o}} \right).
\end{align}
From the above expressions one can see  that when $(b_o=1)$ the spacetime is vacuum, and when $b_o \not= 1$, an effective matter field exists, which always violates the weak energy condition \cite{Hawking:1973uf}.

 \subsubsection{Spacetime properties near the BH horizon} \label{sec3A1}

To study the properties of the spacetime near the BH horizon, an important quantity is the Hawking temperature. Following \cite{Ashtekar:2020ckv}, for a spherically symmetric static spacetime of the form
\begin{align}
    \label{eq3.7a}
    ds^2 = - g_{tt}  dt^2 + g_{xx} dx^2 + p_c d\Omega^2,
\end{align}
the Hawking temperature is given by
\begin{align}
    \label{eq3.7b}
    T_{H}=\frac{\hbar}{k_B \mathcal{P}}, \quad
\mathcal{P} = \lim_{t\to 0} \frac{4 \pi \left(g_{tt}g_{xx}\right)^{\frac{1}{2}}}{\partial_{t} g_{xx}},
\end{align}
where $k_B$ is the Boltzmann constant. We find that, for the metric (\ref{eq3.3}), the  Hawking temperature is given by
\begin{align}
\label{eq3.7}
    T_{H} = \frac{T_H^{\text{GR}}}{\cal{P}}, 
\end{align}
where $T_H^{\text{GR}} \left(\equiv \hbar/(8\pi k_b m)\right)$ denotes the corresponding  Hawking temperature of the classical BH and
\begin{align}
\label{eq3.8}
    {\cal{P}} \equiv \left(1+c_o^2\right)\frac{\gamma L_o \delta_c}{8 m c_o}.
\end{align}
It is clear that we  must choose
\bqn
\lb{co}
c_o = \frac{\gamma L_o \delta_c}{8 m},
\eqn
in order to make sure that the temperatures of massive BHs do not deviate significantly from their classical counterparts.
It is remarkable to note that  the above choice is precisely the one adopted by   CS and AOS \cite{Corichi:2015xia,Ashtekar:2018cay}, given by Eq.(\ref{eq2.23}), for which we have
 \begin{align}
\label{eq3.8b}
    {\cal{P}}^{\text{(CS, AOS)}}  =  1+ c_o^2.
\end{align}

\subsection{Spacetime inside the BH/WH horizons} \label{sec3B}

Recall that the metric inside the BH/WH horizons for the BH takes the form
\begin{align}
\label{eq3.19}
ds^2 = - {N}^2 d{T}^2 + {g}_{xx} dx^2 + {p}_cd\Omega^2,
\end{align}
where  $N$, ${g}_{xx}$ and $p_c$ are given by Eq.(\ref{eq2.31}).
In the rest of this paper, we shall adopt the value of $c_o$ given by Eq.(\ref{co}),
which will ensure the quantum corrections to the Hawking temperature are negligible for macroscopic BHs. This choice reduces the four-dimensional phase space to three, {now spanned by} ($m, \delta_b, \delta_c$). However, unlike previous studies, we shall keep the choices of  $\delta_b$ and $\delta_c$ open.

To gain a deeper understanding of this three-dimensional phase space, let us first consider the spacetime near the transition surface.

\subsubsection{Properties of the spacetime near the transition surface} \label{sec3B1}

To our goals, let us focus ourselves on the behavior of the Kretschmann scalar
  near the transition surface. Since the expression of the Kretschmann scalar is humongous, its explicit form will not be written out explicitly in this paper.

To begin with, let us first note that the Kretschmann scalar is independent of the mass parameter $m$ at the  transition surface for the AOS choice of ($\delta_b, \delta_c$),
given by Eq. (\ref{eq1.2}) \cite{Ashtekar:2018cay}. Therefore, we would first
consider the case in which $\delta_b$ and $\delta_c$ take the forms
\begin{align}
    \label{eq3.21new}
    \delta_b = \alpha_b \left(\frac{\ell_p}{m}\right)^{1/3}, \quad L_o \delta_c = \alpha_c \left(\frac{\ell_p}{m}\right)^{1/3},
\end{align}
where $\alpha_b$ and $\alpha_c$ are two positive and otherwise arbitrary constants. For the AOS solution, we have
\bqn
\lb{alphas_AOS}
  \alpha^{\text{(AOS)}}_b &=& \left(\frac{\sqrt{\Delta}}{\sqrt{2\pi} \gamma^2 \ell_p}\right)^{1/3}, \nb\\
     \alpha^{\text{(AOS)}}_c &=& \frac{1}{2}\left(\frac{\gamma\Delta^2}{4\pi^2 \ell_p}\right)^{1/3}.
\eqn

However, we find that for any {nonzero and positive} values of ($\alpha_b, \alpha_c$), the Kretschmann scalar $K_{\cal{T}}$ evaluated at the transition surface is always independent of the mass parameter $m$. In Fig. \ref{Fig1}, we plot three different choices,
    $\left(\alpha_b,\alpha_c\right)=\left\{(1.97,3),\; (2,0.5),\; (1,1)\right\}$,
denoted by lines a, b, and c, respectively, from which one can see clearly that $K_{\cal{T}}$ is independent of mass. To compare them with the case considered in \cite{Ashtekar:2018lag,Ashtekar:2018cay}, the AOS choice of Eq.(\ref{alphas_AOS}) is also plot out and denoted by the line AOS. It should be also noted that in each of these cases $K_{\cal{T}}$ does not depend on $m$, but it does depend on the specific values of ($\alpha_b, \alpha_c$).

\begin{figure}[htbp]
\includegraphics[width=0.95\linewidth]{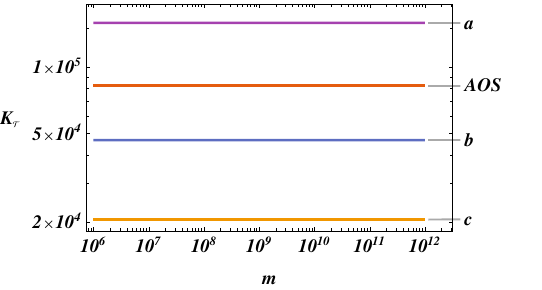}
\caption{ The Kretschmann scalar $K_{\cal{T}}$ evaluated at the transition surface   vs
the mass parameter $m$ for different choices of ($\delta_b, \delta_c$). Lines a, b, and c correspond,  respectively, to the choices  $(\alpha_b,\alpha_c)=\{(1.97,3);\; (2,0.5);\; (1,1)\}$.
The line AOS corresponds to the choice of Eq. (\ref{alphas_AOS}).}
\label{Fig1}
\end{figure}

To see how $K_{\cal{T}}$ depends on the powers of $m$ appearing in $\delta_b$ and $\delta_c$, let us consider the case
\bqn
    \label{eq3.22}
    \delta_b = \alpha_b \left(\frac{\ell_p}{m}\right)^{\beta_b}, \quad L_o \delta_c = \alpha_c \left(\frac{\ell_p}{m}\right)^{\beta_c},
\eqn
where $\beta_b$ and $\beta_c$ are other two arbitrary constants. In particular, in Fig. \ref{Fig2} we choose
\bqn
\lb{eq3.23aa}
\left(\delta_b, \delta_c\right) = \left(\delta^{\text{(AOS)}}_b, \frac{1}{{2 L_o}}{\left(\frac{\gamma  \Delta ^{3/2}}{4 \pi ^2 m}\right)^{1/2}}\right).
\eqn
From this figure we can see that now $K_{\cal{T}}$ increases as $m$ is {increasing}.
On the other hand, in Fig. \ref{Fig3}, we consider the case
\bqn
\lb{eq3.23bb}
\left(\delta_b, \delta_c\right) = \left(\left(\frac{\sqrt{\Delta }}{\sqrt{2 \pi } \gamma ^2 m}\right)^{1/2}, \delta^{\text{(AOS)}}_c\right),
\eqn
from which we can see that $K_{\cal{T}}$ is also increasing as $m$ increases. 

 In addition, if $\beta_{b, c} < 1/3$, then  $K_{\cal{T}}$ will be decreasing as $m$ increases and when $\beta_{b, c} > 1/3$,   $K_{\cal{T}}$ increases as $m$ increases. To show these properties,  in Fig. \ref{Fig4}, we plot the case
 \bqn
 \lb{eq3.23cc}
 \left(\delta_b, \delta_c\right) = \left(\left(\frac{\sqrt{\Delta }}{\sqrt{2 \pi } \gamma ^2 m}\right)^{1/2}, \frac{1}{{2 L_o}}{\left(\frac{\gamma  \Delta ^{3/2}}{4 \pi ^2 m}\right)^{1/2}}\right), ~~~~~
 \eqn
 denoted by line f, and the one
 \bqn
 \lb{eq3.23dd}
 \left(\delta_b, \delta_c\right) = \left(\left(\frac{\sqrt{\Delta }}{\sqrt{2 \pi } \gamma ^2 m}\right)^{1/4}, \frac{1}{{2 L_o}}{\left(\frac{\gamma  \Delta ^{5/2}}{4 \pi ^2 m}\right)^{1/4}}\right), ~~~~~
 \eqn
 denoted by line g.

\begin{figure}[htbp]
\includegraphics[width=0.95\linewidth]{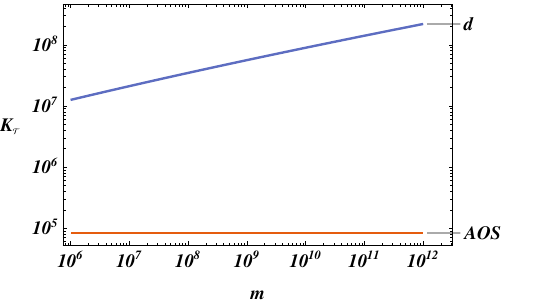}
\caption{Plot of $K_{\cal{T}}$ vs $m$ for the choice of ($\delta_b, \delta_c$) given by Eq.(\ref{eq3.23aa}) with $\left(\beta_b, \beta_c\right) = \left(1/3, 1/2\right)$.  To compare with the AOS choice, Eq.(\ref{alphas_AOS}), the corresponding line is also plotted out, denoted by AOS.}
\label{Fig2}
\end{figure}

\begin{figure}[htbp]
\includegraphics[width=0.95\linewidth]{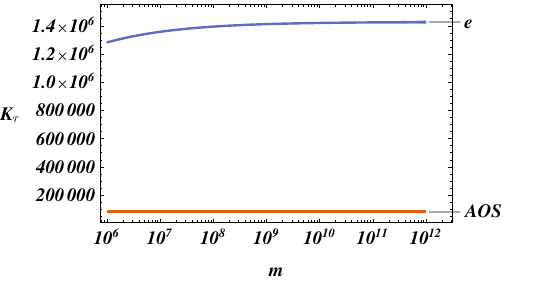}
\caption{Plot of $K_{\cal{T}}$ vs $m$ for the choice of  Eq.(\ref{eq3.23bb}) with $\left(\beta_b, \beta_c\right) = \left(1/2, 1/3\right)$.  The AOS choice, Eq.(\ref{alphas_AOS}),  corresponds to the line AOS.}
\label{Fig3}
\end{figure}

\begin{figure}[htbp]
\includegraphics[width=0.95\linewidth]{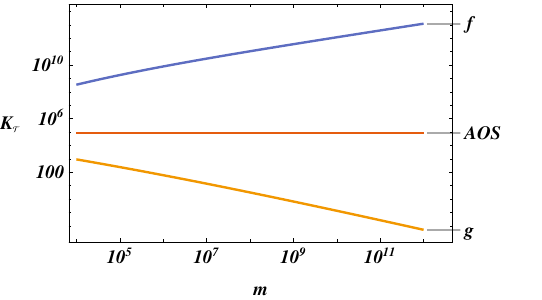}
\caption{Plot of $K_{\cal{T}}$ vs $m$. Line f corresponds to the choice of Eq.(\ref{eq3.23cc})
with $\left(\beta_b, \beta_c\right) = \left(1/2, 1/2\right)$ and line g to the choice of Eq. 
 (\ref{eq3.23dd}) with $\left(\beta_b, \beta_c\right) = \left(1/4, 1/4\right)$.}
\label{Fig4}
\end{figure}

From these figures it is  clear that $K_{\cal{T}}$ will be independent of the mass parameter $m$, as long as $\delta_{b, c}$'s are all proportional to $m^{-1/3}$, no matter what the coefficients $\alpha_{b,c}$'s will be. The AOS choice is only a point on the two-dimensional plane spanned by $\left(\alpha_{b}, \alpha_c\right)$, given by Eq.(\ref{alphas_AOS}).  At any point of this plane, $K_{\cal{T}}$ will not depend on the choice
of $m$. On the other hand, outside of this plane, that is, as long as $\beta_b \not = 1/3$ and/or $\beta_c \not = 1/3$, $K_{\cal{T}}$ will depend on $m$.

In \cite{Ashtekar:2018lag,Ashtekar:2018cay}, $\delta_{b}$ and $\delta_c$  were uniquely determined by requiring that the physical areas of the plaquettes $\Box(\theta, \phi)$ and $\Box(x, \phi)$ at the transition surface be equal to the area gap $\Delta$. {A natural question to ask is whether there exist other conditions that can uniquely determine}  $\delta_{b}$ and $\delta_c$. Our above considerations show that requiring that $K_{\cal{T}}$ be independent of the mass parameter $m$ only uniquely fixed $\beta_b$ and $\beta_c$, but not their amplitudes
 $\alpha_b$ and $\alpha_c$.  If we further require that the masses of the BH and WH be equal,
 can we uniquely determine the amplitudes $\alpha_b$ and $\alpha_c$? To answer these questions, let us turn to consider the radii of the BH and WH horizons.

\subsubsection{Properties of spacetime near WH/BH horizons} \label{sec3B2}

The geometric radii of the WH and BH horizons are given by $r_{A} = \sqrt{p_c(T_A)}$, where
$T_A$ denotes the location of the horizon $A$. In Fig. \ref{Fig5}, we plot the ratio between the two radii for the choices of $\delta_{b}$ and $\delta_c$, given by Eqs. (\ref{eq3.23cc}) and (\ref{eq3.23dd}), denoted, respectively, by line x and y. Line AOS corresponds to the AOS choice.

\begin{figure}[htbp]
\includegraphics[width=0.95\linewidth]{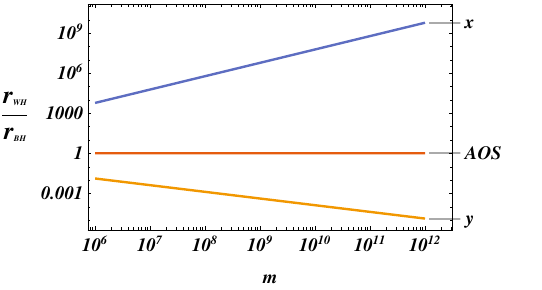}
\caption{Plots of the ratio $r_{\text{WH}}/r_{\text{BH}}$ vs the mass parameter $m$. The lines
$x$,  AOS, and $y$ correspond to the choices of  ($\delta_b, \delta_c$), given, respectively, by
Eqs. (\ref{eq3.23cc}),  (\ref{eq1.2}), and (\ref{eq3.23dd}).}
\label{Fig5}
\end{figure}

On the other hand, from Eqs. (\ref{eq2.20}) and (\ref{eq2.31}), we find that $r_{\text{BH}} = r_{\text{WH}}$ implies that
\begin{align}
\label{eq3.21}
   L_o \delta_c 
    &= \frac{8 m}{\gamma} \left(\frac{\sqrt{\gamma ^2 \delta _b^2+1}-1}{\sqrt{\gamma ^2 \delta _b^2+1}+1}\right)^{{2}/{\sqrt{\gamma ^2 \delta _b^2+1}}}.
\end{align}
Substituting Eq.(\ref{eq3.21new}) into the above equation, we find that { for massive BHs}
\begin{align}
    \label{eq3.22new}
    \alpha_c =   \left(\frac{\gamma^3\ell_p}{2}\right)\alpha_b^4.
\end{align}
Therefore, by imposing the conditions that (a)  $K_{\cal{T}}$ be independent of $m$, and (b) $r_{\text{BH}} = r_{\text{WH}}$, the two parameters $\delta_b$ and $\delta_c$ can be uniquely determined by Eqs. (\ref{eq3.21new}) and (\ref{eq3.22new}), additionally yielding a free parameter $\alpha_b$.
In other words, there exits a family of choices of $\delta_b$ and $\delta_c$, for which the above  kinds of properties are true.

\section{Conclusions} \label{sec4}
\renewcommand{\theequation}{4.\arabic{equation}}
\setcounter{equation}{0}

In this paper, we have systematically studied the most general solutions  of LQBHs of
 the effective Hamiltonian obtained from its classical counterpart by the replacements (\ref{eq1.1}), with $\delta_b$ and $\delta_c$ being arbitrary constants. These solutions usually contain five free parameters: three are the integration constants of the dynamical equations \cite{ElizagaNavascues:2022npm}, and two are the polymerization parameters $\delta_b$ and $\delta_c$. However, using the gauge freedom (\ref{GTs}),   one of them can be gauged away simply by the replacement, $T \rightarrow T + \hat T_o$, as shown explicitly
 by Eqs. (\ref{Replacements})-(\ref{eq2.19}). Therefore, there are generically only four free parameters, $c_o$, $m$, $\delta_b$, and $\delta_c$. These solutions were already studied by several authors from different aspects, including Refs. \cite{ElizagaNavascues:2022npm,Corichi:2015xia,Ashtekar:2018cay}.

 However, our current studies are different from those previous ones in the sense that we have also taken $\delta_b$ and $\delta_c$ as free parameters and explored their effects on various properties of the LQBH spacetimes. On the other hand, in \cite{Corichi:2015xia} Corichi and Singh studied the solutions with the choice (\ref{Deltas_CS}), while in
 \cite{Ashtekar:2018cay,ElizagaNavascues:2022npm} the choice (\ref{eq1.2}) was adopted.

The four-parameter solutions are usually obtained in the KS spacetime, as shown explicitly by Eqs.(\ref{metricB})  and (\ref{MCsA}), which becomes singular at both of the WH and BH horizons.  Therefore, extensions beyond these horizons are needed, in order to study the asymptotic properties of the spacetimes. We have carried out such analytical extensions in Sec. \ref{sec2C}. Since the extensions are analytical, they are also unique. Once these are done, we are allowed to study the spacetimes in each of the three regions separately: the two asymptotic regions outside of the WH and BH horizons and the internal region in between them.

Since the two asymptotic regions have quite similar properties, it is sufficient to study only one of them, which we choose the one on the BH side with $T > 0$ (cf. Sec. \ref{sec3A2}). For any given four free parameters ($c_o, m, \delta_b, \delta_c$), the Kretschmann scalar in the asymptotic limit always takes the form of Eqs. (\ref{eq3.9}) and (\ref{eq3.9aa}). From these expressions, it can be seen that the leading order is $1/r^4$, as long as $\delta_b \not= 0$, irrespective of any specific values of the four parameters. On the other hand, when $\delta_b = 0$, for which we have $b_o = 1$, it reduces to its classical value, $K_{\text{GR}} = 48m^2/r^6$. {However, more careful analysis showed that the $r^{-4}$ term becomes dominant over the $r^{-6}$ term only when $r > r_{c}$, where $r_{c}$ is defined by Eq.(\ref{eq3.10a}). In addition, we also show that the $r^{-4}$ term dominates over the $r^{-5}$ term only when $r > r_{c_4}$, given in Eq. (\ref{eq3.15}). And we further show that $r^{-5}$ is dominant over the $r^{-6}$ term only when $r > r_{c_5}$, also given in Eq. (\ref{eq3.15}). For both the AOS and CS choices, we found that $r_c$, $r_{c_4}$, and $r_{c_5}$  are greater than $L_{\text{obs}}$ for solar mass BHs, where $L_{\text{obs}}\; (\simeq 4.4 \times 10^{28} \; {\text{cm}})$ denotes the  size of our observational Universe. Therefore, in both cases,  the asymptotic behavior of the spacetime can be well described by its classical limit ($K \simeq r^{-6}$) within our observational Universe.}

{We have further calculated the Hawking temperature of the BH and found that it is given by Eqs. (\ref{eq3.7}) and (\ref{eq3.8})}. For macroscopic BHs, the gravitational fields near the BH horizons are very weak, and we expect that the  temperature should be very close to that of classical BHs. With this requirement, it is clear that the free parameter $c_o$ must be chosen so that Eq.(\ref{co}) holds, i.e.,
\bqn
\lb{eq4.1}
c_o = \frac{\gamma L_o \delta_c}{8 m}.
\eqn
It is remarkable to note that this was precisely the choice adopted in
\cite{Corichi:2015xia,Ashtekar:2018cay}.

To explore the effects of the polymerization parameters $\delta_b$ and $\delta_c$, we have first studied the Kretschmann scalar $ K_{\cal{T}}$ at the transition surface $T = T_{\cal{T}}$ and found that it is independent of the mass parameter $m$ as long as they take the forms given by Eq. (\ref{eq3.21new}) for any chosen values of $\alpha_b$ and $\alpha_c$, as shown explicitly in Fig. \ref{Fig1}. Clearly, the AOS choice of Eq. (\ref{alphas_AOS}) is only a point in the
plane spanned by ($\alpha_b, \alpha_c$). To see how $ K_{\cal{T}}$ depends on the powers of $m$ in the expressions of $\delta_b$ and $\delta_c$, we have studied the more general forms, Eq. (\ref{eq3.22}), and found that $\beta_b = \beta_c = 1/3$ is the unique choice that leads $ K_{\cal{T}}$ to be independent of $m$, as shown explicitly by Figs. \ref{Fig2}-\ref{Fig4}.

Finally, we have also explored the condition at which the WH and BH masses are equal
 and found it is given by Eq. (\ref{eq3.21}). If we further require that the Kretschmann scalar   at the transition surface be independent of $m$, we find that  $\delta_b$ and $\delta_c$
 must take the form (\ref{eq3.21new}) with $\alpha_c$ being given by Eq. (\ref{eq3.22new}).
 Therefore, it is concluded that {\em as long as $\delta_b$ and $\delta_c$ are chosen as}
 \bqn
 \lb{eq4.2}
    \delta_b = \alpha_b \left(\frac{\ell_p}{m}\right)^{1/3}, \;\;
    L_o \delta_c = \alpha_b^4 \left(\frac{\gamma^3\ell_p}{2}\right) \left(\frac{\ell_p}{m}\right)^{1/3}, ~~~~
\eqn
 {\em the corresponding LQBHs shall have the same desired properties as those of the AOS solution \cite{Ashtekar:2018cay}, for any given {nonzero and positive} $\alpha_b$.}

{With the choice of $\delta_b$ and $\delta_c$ given by Eq.(\ref{eq4.2}), an interesting question is whether it is possible to impose any constraint on the choice of the free parameter $\alpha_b$ from observations. As the  Event Horizon 
Telescope just observed two supermassive BHs, one is 
 the $M87^*$ BH located at the center of the more distant Messier 87 galaxy \cite{EventHorizonTelescope:2019dse}, and the  other is the Sagittarius $A^*$  BH located at the center of our own Milky Way Galaxy \cite{EventHorizonTelescope:2022wkp}, it would be very interesting to see the effects of $\alpha_b$ on the shadows of supermassive BHs. In Fig. \ref{Fig6}, we plot the dependence of $r_c$
 on $\alpha_b$ for a solar mass and Sagittarius $A^*$ and $M87^*$ BHs,
 respectively, from which it can be seen that $r_c$ becomes very small, so that the quantum effects are expected  to be large, only when $\alpha_b$ is very large.  For more details, we would like to come back to this issue on another occasion, as such considerations clearly are out of the scope of the current paper.}

\begin{figure}[htbp]
\includegraphics[width=0.95\linewidth]{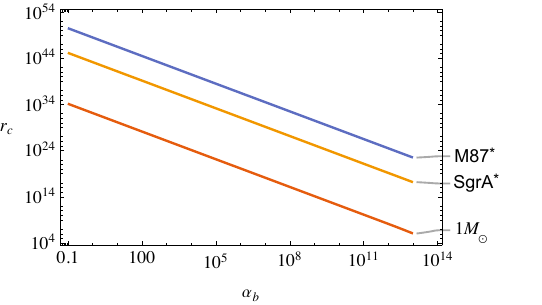}
\caption{Plots of the critical radius $r_c$ vs $\alpha_b$ for a solar mass and Sagittarius $A^*$ and $M87^*$ BHs,
respectively.}
\label{Fig6}
\end{figure}

\section*{ACKNOWLEDGMENTS}
\vspace{-1em}
G.O. is supported through Baylor Physics graduate program, P.S. is supported by NSF Grant  No. PHY-2110207, and A.W. is partly supported by the NSF Grant No. PHY-2308845.

\bibliographystyle{apsrev4-1}
\bibliography{main}
\end{document}